\documentclass[runningheads]{llncs}
\usepackage[height=235mm,width=155mm,center]{crop}

\usepackage{longtable}

\usepackage{eurosym}

\usepackage[caption=false]{subfig}

\usepackage[width=0.5cm]{progressbar}

\usepackage{wasysym}

\usepackage[para,online,flushleft]{threeparttable}

\usepackage[inline]{enumitem}

\usepackage{booktabs, hhline}
\usepackage{xcolor,colortbl}
\definecolor{Gray}{gray}{0.98}

\setenumerate{label=(\roman*), leftmargin=2em}

\newlist{answerlist}{enumerate}{1}
\setlist*[answerlist,1]{%
	label=\Circle,
}

\newcommand{\censorhidden}[1]{}
\newcommand{\censorchange}[2]{#2}
\newcommand{\censor}[1]{\textit{\textless withheld during blind review \textgreater}}

\newcommand{\website}{\emph{[website]}~}

\renewcommand{\censorhidden}[1]{#1}
\renewcommand{\censorchange}[2]{#1}
\renewcommand{\censor}[1]{#1}

\usepackage{soul}
\usepackage{xcolor}

\definecolor{hlcolor}{RGB}{214, 239, 255}\sethlcolor{hlcolor}%

\let\llncssubparagraph\subparagraph
\let\subparagraph\paragraph
\usepackage[compact]{titlesec}
\let\subparagraph\llncssubparagraph

\titlespacing*{\section}       {0pt}{*4}{*1}
\titlespacing*{\subsection}    {0pt}{*2}{*0.5}
\titlespacing*{\section}       {0pt}{*2}{*1}
\titlespacing*{\subsection}    {0pt}{*2}{*1}
\titlespacing*{\subsubsection} {0pt}{*0.9}{*0.9}
\usepackage[hidelinks,implicit=true,bookmarksdepth=2]{hyperref}

\hypersetup{
	colorlinks=true,
	citecolor=blue,
	linkcolor=blue,
	urlcolor=blue,
	pdftitle={What's in Score for Website Users: A Data-driven Long-term Study on Risk-based Authentication Characteristics (Pre-proceedings Version)},
	pdfauthor={Stephan Wiefling,Markus Dürmuth,Luigi Lo Iacono},
	pdfsubject={Risk-based authentication (RBA) aims to strengthen password-based authentication rather than replacing it. RBA does this by monitoring and recording additional features during the login process. If feature values at login time differ significantly from those observed before, RBA requests an additional proof of identification. Although RBA is recommended in the NIST digital identity guidelines, it has so far been used almost exclusively by major online services. This is partly due to a lack of open knowledge and implementations that would allow any service provider to roll out RBA protection to its users. To close this gap, we provide a first in-depth analysis of RBA characteristics in a practical deployment. We observed N=780 users with 247 unique features on a real-world online service for over 1.8 years. Based on our collected data set, we provide (i) a behavior analysis of two RBA implementations that were apparently used by major online services in the wild, (ii) a benchmark of the features to extract a subset that is most suitable for RBA use, (iii) a new feature that has not been used in RBA before, and (iv) factors which have a significant effect on RBA performance. Our results show that RBA needs to be carefully tailored to each online service, as even small configuration adjustments can greatly impact RBA's security and usability properties. We provide insights on the selection of features, their weightings, and the risk classification in order to benefit from RBA after a minimum number of login attempts.},
	pdfkeywords={Risk-based Authentication (RBA),Authentication features,Big Data Analysis,Usable Security}
}

\usepackage{marvosym}
\usepackage{./orcidlink}

\renewcommand{\orcidID}[1]{\footnotesize\orcidlink{#1}\normalsize}

\newcommand{\correspondingauthor}{\textsuperscript{\small(\Letter)}}

\usepackage{textpos}
\newcommand{\copyrightnotice}{
\begin{textblock}{9}(0.0,8.35)
	\noindent
	\scriptsize Pre-proceedings version of a paper accepted for FC 2021.\\The post-proceedings version will be available at Springer after the conference.\\
	\url{https://link.springer.com/conference/fc}
\end{textblock}
\vspace{-1.3em}%
} %

\begin{document}
	\title{What's in Score for Website Users: \\A Data-driven Long-term Study on \\Risk-based Authentication Characteristics}
	\titlerunning{A Data-driven Long-term Study on RBA Characteristics}
	\author{Stephan Wiefling\inst{1,2}\correspondingauthor\orcidID{0000-0001-7917-6065} 
	    \and
		Markus D\"urmuth\inst{2} \and
		Luigi Lo Iacono\inst{1}\orcidID{0000-0002-7863-0622}
	}
	\authorrunning{S. Wiefling et al.}
	\institute{H-BRS University of Applied Sciences, Sankt Augustin Germany \\
		\email{\{stephan.wiefling,luigi.lo\_iacono\}@h-brs.de} \and
		Ruhr University Bochum, Bochum, Germany \\
		\email{\{stephan.wiefling,markus.duermuth\}@rub.de}
	}
	\maketitle              %
	\copyrightnotice
	\vspace{-1em}
	\begin{abstract}
Risk-based authentication (RBA) %
aims to strengthen pass\-word-based authentication rather than replacing it. RBA does this by monitoring and recording additional features during the login process. If feature values at login time differ significantly from those observed before, RBA requests an additional proof of identification. Although RBA is recommended in the NIST digital identity guidelines, it has so far been used almost exclusively by major online services. This is partly due to a lack of open knowledge and implementations that would allow any service provider to roll out RBA protection to its users.

To close this gap, we provide a first in-depth analysis of RBA characteristics in a practical deployment. We observed N=780 users with 247 unique features on a real-world online service for over 1.8 years. Based on our collected data set, we provide 
\begin{enumerate*}
    \item a behavior analysis of two RBA implementations that were apparently used by major online services in the wild,
    \item a benchmark of the features to extract a subset that is most suitable for RBA use,
    \item a new feature that has not been used in RBA before, and
    \item factors which have a significant effect on RBA performance.
\end{enumerate*}
Our results show that RBA needs to be carefully tailored to each online service, as even small configuration adjustments can greatly impact RBA's security and usability properties. We provide insights on the selection of features, their weightings, and the risk classification in order to benefit from RBA after a minimum number of login attempts. 		\keywords{Risk-based Authentication (RBA) 
		\and Authentication features
		\and Big Data Analysis 
		\and Usable Security.}
	\end{abstract}
\section{Introduction}\label{sec:intro}

Despite their long known weaknesses~\cite{morris_password_1979,bonneau_science_2012,wang_targeted_2016,von_zezschwitz_honey_2014,das_tangled_2014,florencio_large-scale_2007,dhamija_why_2006}, passwords are still used for authentication on most online services~\cite{quermann_state_2018}. However, threats to password-based authentication continue to evolve to attacks involving targeted guessing~\cite{wang_targeted_2016,pal_beyond_2019} or stolen credentials sourced from data breaches~\cite{thomas_protecting_2019}.

Thus, online services need to implement alternative or additional measures to protect their user base. Two-factor authentication (2FA) is such a measure, but tends to be only accepted in online banking use cases~\cite{reynolds_tale_2018,dutson_dont_2019,wiefling_more_2020}.
Also, universal second factor (U2F) or biometric authentication require additional hardware and active user enrollment, which makes them impractical for online services~\cite{das_why_2018,gaddam_usage_2019}.

For these reasons, several major online services deployed risk-based authentication (RBA) to protect their users~\cite{wiefling_is_2019}. RBA is an adaptive authentication mechanism which increases password security with minimal impact on the user. It achieves better usability than comparable 2FA methods~\cite{wiefling_more_2020} and is recommended by NIST~\cite{grassi_digital_2017} to mitigate credential stuffing.

During the password entry, RBA monitors and records features that are available in this context. These feature range from network information%
, or device information%
, to behavioral information%
. Based on these features, RBA calculates a risk score related to the login attempt%
. The score is typically classified by an access threshold into low, medium, and high risk~\cite{freeman_who_2016,molloy_risk-based_2012,hurkala_architecture_2014}. Based on the estimated risk, the RBA system can invoke multiple actions. If the score is under the %
threshold, i.e., a low risk%
, access is granted. If the score is above this threshold, i.e., medium or high risk%
, the online service asks for additional information (e.g., confirming an email address) or even blocks access.

RBA schemes, their configuration, and features have not been researched thus far. These are, however, of crucial importance, since they can highly impact security and usability for website users. A feature might reduce the number of re-authentication requests but could also weaken the attack protection%
. %
To further investigate this topic, we formulated the following research questions.

\subsubsection{Research Questions.}

With these research questions, we aim to to provide answers on how RBA performs in a practical deployment and how RBA can be configured to provide the best balance between security and usability.

\newlist{RQLIST}{enumerate}{1}
\setlist[RQLIST]{label=\bfseries RQ\arabic*:, leftmargin=3.2em, parsep=0em}

\newlist{RQ2LIST}{enumerate}{2}
\setlist[RQ2LIST]{label=\bgroup\bfseries \alph*)\egroup,leftmargin=1.6em, parsep=0em}

\begin{RQLIST}
    \item \begin{RQ2LIST}
        	\item How often does RBA request for re-authentication in a practical deployment?
        	\item How many user sessions need to be captured and stored in the login history to achieve a stable and reliable RBA setup?
        \end{RQ2LIST}

	\item \begin{RQ2LIST}
	    \item Which RBA features have to be chosen to achieve good security?
	    \item How do RBA features need to be combined to achieve good security?
		\item How often will different RBA feature combinations request legitimate users for re-authentication%
		?
	\end{RQ2LIST}
	
	\item \begin{RQ2LIST}
		\item How practical are different RBA configurations regarding performance?
		\item How scalable and cost-efficient are different RBA configurations?
	\end{RQ2LIST}
\end{RQLIST}

\subsubsection{Contributions.}

We provide the first long-term data-driven analysis of RBA characteristics%
.
\begin{enumerate*}
    \item We monitored and recorded the login behavior and features of 780 users on a real-world online service for over 1.8 years.
    \item We derived two RBA models %
    based on the majority of deployments used in current practice.
    \item We evaluated the two %
    models on our data set and identified features that, in combination, provide %
    good security and usability.
    \item We proposed and tested a new feature that had not yet been seen in the RBA and browser fingerprinting context before.
    \item We derived how specific factors influence RBA's %
    performance.
\end{enumerate*}

The results show that even small changes to RBA settings, e.g., the feature set or access threshold, can strongly affect the usability and security properties of RBA.
Our work supports service owners regarding RBA design decisions on their website. It helps administrators select suitable RBA properties---including the RBA scheme, feature set, and weightings---for their website's characteristics and needs. Finally, researchers obtain insights on RBA's inner workings in practice. Understanding these factors can provide a comprehensive understanding of RBA and foster a widespread %
adoption that goes beyond the current use by only major online services.

\section{RBA Models}\label{sec:rba-solutions}

We derived and evaluated two RBA models %
based %
on observations on the RBA behavior of major %
online services~\cite{wiefling_is_2019} %
and %
algorithm descriptions in literature.%

The \textbf{simple model} (SIMPLE) extends the single-feature model used in the open source single sign-on solution OpenAM~\cite{openam_adaptive_2016} and is assumed to be used at %
GOG.com~\cite{wiefling_is_2019}. It also partly reflects models given in literature~\cite{steinegger_risk-based_2016,hurkala_architecture_2014,djosic_machine_2020}. We based our implementation on %
OpenAM, %
since it is freely available and probably widely used. %
The SIMPLE algorithm checks a number of features for an exact match in the user's login history. The risk score is the number of inspected features with at least one match in the login history divided by the total number of considered features. Thus, the risk score granularity increases with the number of observed features. We tested this model in two variations to observe the potential of OpenAM's original implementation.
For a fair comparison with an influential RBA algorithm in literature~\cite{freeman_who_2016}, the first variation used the features \emph{IP address} with \emph{IP-based geolocation}, and \emph{user agent string} (SIMPLE-IPUA). %
In the second variation, we enabled the maximum %
number of features in the Open\-AM solution %
to test its maximum potential (SIMPLE-ALL). Besides the three features, there were \emph{registered client} (HTML5 canvas and WebGL fingerprint), and \emph{last login} (i.e., logged in within the last 31 days).%

The \textbf{extended model} (EXTEND) %
is comparable to the multi-features model that %
Google, Amazon, and LinkedIn used~\cite{wiefling_is_2019} and presumably still use in some form. We based this model on Freeman et al.~{\cite{freeman_who_2016}}, since it was the only comparable algorithm described in the literature%
. %
The model calculates the risk score $S$ for a user $u$ and a given feature set %
$(x^1,..., x^d)$ with $d$ features as~\cite{freeman_who_2016}:
\begin{equation}
    S_{u}(x) = \left( \prod_{k=1}^{d} \frac{p(x^k)%
    }{p(x^k | u, legitimate)%
    } \right) \frac{p(u | attack)}{p(u | legitimate)}
\end{equation}
 $p(x^k)$ is the probability of a feature value in the global login history and \linebreak $p(x^k | u, legitimate)$ is the probability that a legitimate user has this feature value in its own login history. %
 Since we did not collect attack data, we assumed that all users are equally likely to be attacked. Thus, we set $p(u | attack) = \frac{1}{|U|}$, where $U$ is the set of users with $u \in U$. The probability of legitimate logins for the user is based on the proportion of logins, i.e., $p(u | legitimate) = \frac{Number\ of\ user\ logins}{Number\ of\ all\ logins}$. Since the risk score depends on the global login history size, the risk score granularity increases with the number of entries in the global login history.

We smoothed the features with linear interpolation to add probabilities for previously unseen but plausible values~\cite{freeman_who_2016}. We also subdivided some features into subfeatures with individual weightings (IP address $\rightarrow$ autonomous system number (ASN) and country; user agent string $\rightarrow$ browser/OS name and version, and device type, i.e., mobile or desktop). %
Freeman et al. evaluated these features and subfeatures with the help of LinkedIn~\cite{freeman_who_2016}. Thus, these potentially represent a practical RBA feature set, which is why we chose and tested them as a %
baseline.

\section{Data Set}\label{sec:data-set}

We evaluated the RBA models with a data set containing real-world user behavior to identify the model characteristics in a practical deployment.%

\subsubsection{Data Collection.}

We recorded user data from August 2018 to June 2020 on an e-learning website for medical students. During course enrollment, they were registered at the website by the faculty staff. The students used this online service to exercise for their study courses and exams. %
After each successful login, we collected 247 different features of the user's online browser, network, and device (see Table~\ref{tab:feature-overview} in Appendix~\ref{appendix:features})%
. %
The features were relevant in the field of device fingerprinting~\cite{pugliese_long-term_2020,alaca_device_2016} and could help to identify users in RBA as well.

The data set is very challenging for RBA since the users are mostly located in %
the same city%
. Thus, they could get similar feature values, e.g., IP addresses, with higher probability%
. Testing this data %
will answer %
whether practical RBA deployments can protect users in such a challenging scenario.%

\subsubsection{Survey.}
The e-learning website collected usernames, hashed passwords, and features only%
. After the collection 
phase, we surveyed %
users between July and August 2020 to improve data quality (see Appendix~\ref{appendix:survey} for the questionnaire).

We recruited via a mailing list of \censorchange{the University of Cologne}{a University\footnote{\label{footnote:orgname-omitted}Organization name omitted during blind review}}, addressing students who potentially used the e-learning website between August 2018 and June 2020. We introduced the study as a survey on the overall website perception. %
We drew 12 Amazon vouchers worth \euro 10 among all %
participants after the study.

After %
verifying their account%
, the users were redirected to the survey. Besides demographics, we included some questions about the website experience to distract from our actual study purpose. To improve data quality, we asked whether the users knew about someone illegitimately logging into their website account. We based this question on Shay et al.~\cite{shay_my_2014}. %

\subsubsection{Demographics.} In total, 182 website users (26.6\% of login sessions) answered the survey. 168 users passed the attention check. The users were 61.3\% female and 38.1\% male (0.6\% did not state the gender). The majority of users (79.7\%) were between 18 and 24 years old. The remaining users were 25-34 years (18.5\%), and 35-54 years old (1.8\%). The age and gender distribution corresponds to the expected demographics for such a study course.

\subsubsection{Login Sessions.}

\begin{figure}[t]
	\centering
	\includegraphics[width=0.87\linewidth]{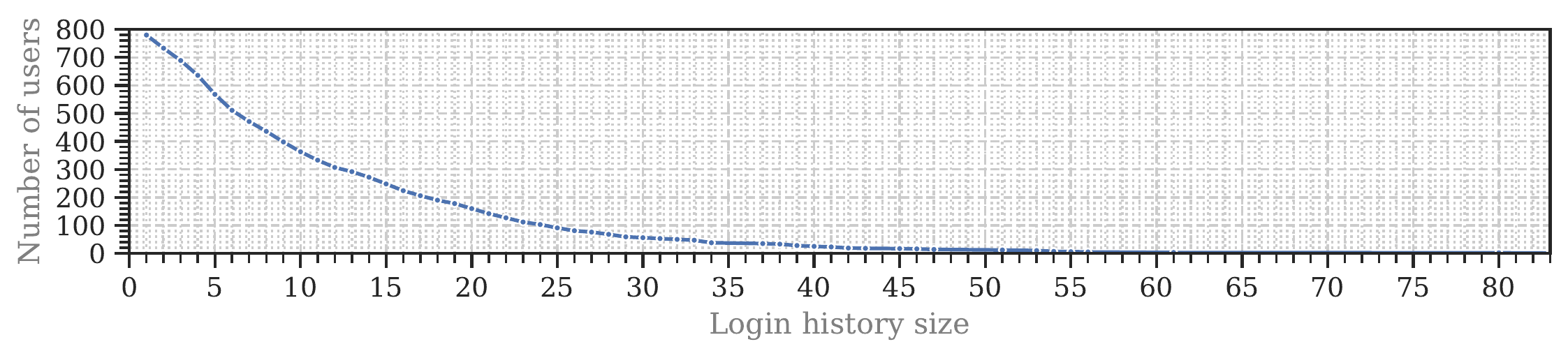}
	\caption{Login history sizes and number of users in our data set%
	}
	\vspace{-1em}
	\label{fig:loginhistorysizes}
\end{figure}

The data set consisted of 780 users and 9555 logins%
. The users mostly logged in daily (44.3\%) or several times a week (39.2\%)%
. They logged in between one and 83 times (mean:~12.25, median:~9, SD:~11.18; see Figure~\ref{fig:loginhistorysizes}). They used desktop (81.1\%) and mobile devices (18.9\%). The desktop devices were Windows (62.5\%), macOS (37.2\%), and Linux (0.3\%) based. Mobile devices were iOS (75.2\%) and Android (24.8\%) based. The browsers were mainly Safari (40.4\%), Chrome (29.0\%), Firefox (26.1\%), and Edge (3.3\%).
To improve the quality and validity of our results, %
we removed users who stated an illegitimate login attempt in the survey. However, there were no such users (93.5\% did not notice, 6.5\% did not know).

\subsubsection{Feature Optimization.}

To improve the expected performance of some of the features, we optimized them based on procedures found in literature~\cite{hurkala_architecture_2014,freeman_who_2016,steinegger_risk-based_2016,alaca_device_2016} and as described in the following.

We extracted additional subfeatures from the IP address, user agent string, and timestamp features. Besides only extracting the hour~\cite{hurkala_architecture_2014}, we also extracted combinations of weekday and hour to gain more information.

Administrators aiming to deploy the EXTEND model need to adjust the feature weightings to appropriate values. Freeman et al.~\cite{freeman_who_2016} did not provide subfeature weightings for IP address and user agent string. Thus, we calculated weightings %
for our data set %
following the method described in their paper. %
As a result, we set the weightings for the IP address (IP address: 0.6, ASN: 0.3, country: 0.1) and user agent (full string: 0.53, browser: 0.27, OS: 0.19, device type: 0.01).
We chose the weightings based on the value of information when present. They only relate to our specific data set, but can give an impression of their distribution in practice.%

\subsubsection{New Feature: Round-Trip Time.}

We propose a new feature that has not been seen in RBA and browser fingerprinting literature at the time of study. In concurrent and independent work, Rivera et al.~\cite{rivera_risk_2020} proposed a similar idea based on the work-in-progress resource timing API. Apart from it being a different approach, their feature is also client originated and thus less trustworthy than our solution.

The web sockets technology~\cite{melnikov_websocket_2011}, which is present in most online browsers today~\cite{caniuse_websockets_2020}, allows measuring the round-trip-time (RTT). The server requests a data packet from the client and measures the time until the response. %
RTTs can give information on whether the user's device is really located in the indicated region, or whether the location was potentially spoofed, e.g., by VPNs or %
proxies~\cite{abdou_secure_2018,campobasso_impersonation_2020}. %
This is also true in the presence of Content Delivery Networks (CDNs), where the CDN edge node can be linked to the RTT. This results in an even better measurement, since the edge nodes %
close to the user's device are also considered.%

When users entered the login credentials, we measured the RTT %
five times%
. %
Then, we stored the smallest RTT value %
to get the best possible %
value and to mitigate larger RTT variations, e.g., due to mobile connectivity. Besides the RTT in microseconds (RTT-RAW), we stored RTTs in milliseconds (RTT-MS), and rounded to the nearest five (RTT-5MS) and ten~milliseconds (RTT-10MS).

\subsubsection{Legal and Ethical Considerations.}

The participants were part of a model medical education program. During enrollment, they signed a consent form agreeing to the data collection for study purposes. They were always able to view their data on request. The collected data was stored on encrypted hard drives. Only the study researchers had %
access to them. The passwords on the website were hashed with scrypt~\cite{percival_scrypt_2016}. All participants gave informed consent on these procedures%
. All survey questions included a ``don't know'' option.

We do not have a formal %
IRB %
process at our university. But besides our ethical considerations above, we made sure to minimize potential harm by complying with the ethics code of \censorchange{the German Sociological Association (DGS)}{a nationwide sociological association} and the standards of good scientific practice of \censorchange{the German Research Foundation (DFG)}{a nationwide research funding organisation\footnote{\label{footnote:orgname-omitted}Organization names omitted during blind review}}. We also made sure to comply with %
the EU General Data Protection Regulation%
.

\section{Attacker Models}

We evaluated the RBA systems using three attacker models based on known ones in the RBA context~\cite{freeman_who_2016,wiefling_evaluation_2020}. All attackers possess the victim's login credentials%
.

\begin{figure}%
    \vspace{-1em}
    \centering
    \includegraphics[width=0.65\linewidth]{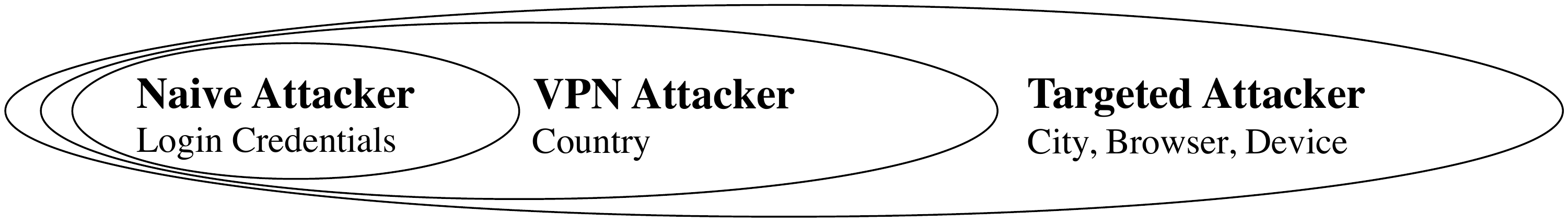}
    \caption{Overview of the attacker models tested in the study}
    \vspace{-1em}
    \label{fig:attacker-models}
\end{figure}

The \textbf{naive attacker} tries to log in via an IP address of a random ISP located somewhere in the world and uses popular user agent strings. %
We simulate these attackers using a random subset of IP addresses sourced from real-world online attacks%
~\cite{firehol_all_2020}. %
Other feature values not related to the IP address are sourced from our data set%
. %
The \textbf{VPN attacker} knows the same as the naive attacker plus the correct country of the victim. The attacker spoofs the IP geolocation with VPN services and uses popular user agent strings. %
We simulate these attackers with known attacker IP addresses~\cite{firehol_all_2020} located in the victim's country. Feature values not derived from the IP address are sourced from our data set. 
We also included IP addresses not directly related to VPN services to consider services that tunnel traffic through client devices%
. %
The \textbf{targeted attacker} extends the knowledge of the VPN attacker by including locations and user agents of the victim. The attacker accesses IP addresses of ISPs in that location, likely including the victim's ones. This attacker is identical to Freeman et al.'s \emph{phishing attacker}~\cite{freeman_who_2016}. We used a different term, however, as phishing is just one of the ways to obtain this level of knowledge.
We simulate this attacker with our data set. The %
feature values are taken from all users except the victim. Since location dependent feature values in our data set were in close proximity to each other%
, our simulated attacker is aware of these circumstances and chooses feature values in a similar way.%

\section{Evaluating RBA Practice (RQ1)} \label{sec:evaluating-rba-practice}

Below, we analyze the RBA behavior in a practical deployment. We describe our methodology to reproduce the RBA behavior and present the results.

\subsubsection{Step 1: Calibrating Risk Scores.} \label{subsec:calibrating-risk-scores}

The risk scores of the RBA models have different granularity %
(see Section~\ref{sec:rba-solutions}). For a fair comparison, we calibrated the risk score access thresholds of both RBA models. We adjusted regarding the percentage of blocked attacks in each attacker model, which we call the true positive rate (TPR), as in related work~\cite{freeman_who_2016}. We approximated the TPRs as close as possible. However, due to their granularity properties, SIMPLE TPRs were more coarse-grained than those of EXTEND.

\subsubsection{Step 2: Determine Re-Authentication Count.}

By replaying user sessions, we determined how often the data set's legitimate users were asked for re-authentication based on the number of logins. %
For each login attempt, we
\begin{enumerate*}
	\item restored the state at the time of the login attempt, %
	\item calculated the risk score with the RBA model, and %
	\item finally applied the calibrated RBA access threshold to the risk score and stored the access decision%
	.
\end{enumerate*}

To provide an average estimation of the RBA behavior, we %
calculated the median re-authentication counts and rates for each login history size. %

\subsection{Results}

\begin{figure}[t]
	\centering
	\includegraphics[width=\linewidth]{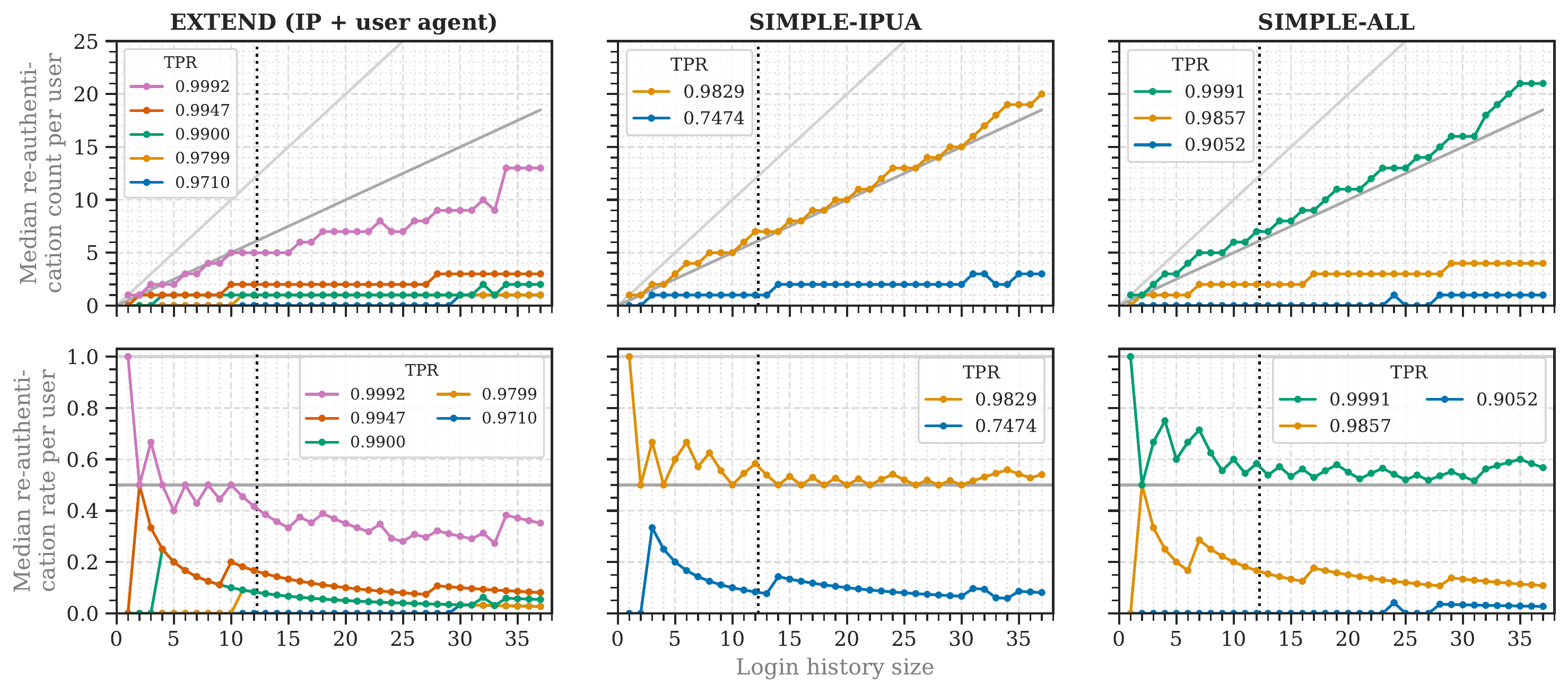}
	\vspace{-2em}
	\caption{Median re-authentication counts (top) and rates (bottom) per user based on the login history size. The TPR (percentage of blocked attacks) relates to targeted attackers. We added the baseline for 2FA (light grey line), the stable setup threshold (dark grey line), and the mean login count (dotted black line) for orientation. Below the stable setup threshold, users had to re-authenticate less than every 2nd login attempt.}
	\vspace{-1em}
	\label{fig:results-reauth-count}
\end{figure}

Figure~\ref{fig:results-reauth-count} shows the results for the targeted attacker case. We answer our research questions regarding practical RBA deployments in the following.%

\subsubsection{Number of re-authentication requests in practice (RQ1a).}
The users logged into the website 12.25 times on mean. Thus, we considered a login history size of 12 to determine the re-authentication count for the average %
user in our data set. We define the median login count until re-authentication as the login history size divided by the median re-authentication count. In the following, we show the results with TPRs adjusted for each attacker model. Note that due to the risk score characteristics, attackers of lower hierarchy were always blocked as well (e.g., all naive attackers were blocked when blocking all VPN attackers).

Even when blocking all \textbf{naive attackers} with the highest possible TPR, legitimate users were never asked for re-authentication at all, except for SIMPLE-IP with TPR 0.999 (every 12th time).
When \textbf{VPN attackers} were blocked, legitimate users were mostly not asked for re-authentica\-tion at all. In the other cases, they were prompted every 2.4th time (TPR 0.9995) and every 12th time (TPR 0.9946, 0.9903) with EXTEND, every 12th time with SIMPLE-IP (TPR 0.9933), and every 6th time with SIMPLE-ALL (TPR 0.9999).
When blocking \textbf{targeted attackers}, our legitimate users were never asked for re-authentication with TPRs lower than 0.98 in most cases (see Table~\ref{tab:login-reauth-count-targeted-attackers} and Figure~\ref{fig:results-reauth-count}). %

\setlength\arrayrulewidth{\heavyrulewidth} 
\begin{table}[t]
    \centering
    \caption{Median login count until re-authentication when blocking targeted attackers}
    \resizebox{0.99\linewidth}{!}{%
    \begin{threeparttable}
        \begin{tabular}{>{\columncolor{Gray}}l@{\hskip 0.5em}rrr@{\hskip 1.5em} >{\columncolor{Gray}}l@{\hskip 0.5em}rrr@{\hskip 1.5em} >{\columncolor{Gray}}l@{\hskip 0.5em}rr}
        \toprule
          &       & Median logins until &       &       &       & Median logins until &       &       &       & Median logins until \\
    Model & TPR   & re-authentication &       & Model & TPR   & re-authentication &       & Model & TPR   & re-authentication \\
\cmidrule{1-3}\cmidrule{5-7}\cmidrule{9-11}    EXTEND & 0.9992 & 2.4   &       & SIMPLE-ALL & 0.9991 & 1.71  &       & SIMPLE-IPUA & 0.9829 & 1.71 \\
          & 0.9947 & 6     &       &       & 0.9857 & 6     &       &       & 0.7474 & 12 \\
          & 0.9900 & 12    &       &       & $<$0.9857 & $\infty$ &       &       & $<$0.7474 & $\infty$ \\
          & 0.9799 & 12    &       &       &       &       &       &       &       &  \\
          & $<$0.9799 & $\infty$ &       &       &       &       &       &       &       &  \\
        \bottomrule
        \end{tabular}%
        \begin{tablenotes}\small
        \item Login history size: 12
        \end{tablenotes}
    \end{threeparttable}
    }
    \vspace{-1em}
    \label{tab:login-reauth-count-targeted-attackers}
\end{table}

Overall, the median re-authentication rate became lower with an increase in the login history size. For very high TPRs, however, the numbers did not decrease to a high degree, especially with the SIMPLE model.

Concluding the results, RBA rarely requests re-authentication for most cases in our real-world data set, even when blocking targeted attackers up to a TPR of over 0.9945 with EXTEND.
However, the re-authentication rate strongly depends on the RBA model and the assumed attacker model. The influence of the feature set and the feature weightings will be analyzed in Section~\ref{sec:features}.

\subsubsection{Required login history size (RQ1b).}

Since RBA is designed to request less re-authentication than 2FA for legitimate users, %
this difference needs to be noticeable in sensible RBA deployments. As a baseline to request less than every second login attempt, we defined the required login history size as the size above which the median re-authentication rate remains below 0.5. For statistical validity, we %
considered login history 
sizes lower than 38 since these had at least 30 users (see Section~\ref{sec:data-set}).

In our data set, most TPRs required one or even no history entry for blocking targeted attackers in both models (see Figure~\ref{fig:results-reauth-count}). However, EXTEND required ten entries %
for TPR 0.9992. The SIMPLE models partly did not fulfill the requirement (TPRs: 0.9829 SIMPLE-IPUA, 0.9991 SIMPLE-ALL). Based on our results, we conclude that storing one entry is already sufficient for a stable setup that blocks more than 99.45\% of targeted attackers with the EXTEND model. To block 99.92\% of attack attempts, ten entries are needed in our use case.

\subsection{Discussion}

Small variations of the access thresholds (see Section~\ref{sec:intro}) can greatly affect the TPR. For instance, changing a tiny fraction of the %
threshold lowered the TPR from a very good 0.9829 to 0.7474 in SIMPLE-IPUA. We assume that this can make it difficult for administrators to adjust the access thresholds correctly. To foster a widespread RBA adoption in the wild, %
we suggest that RBA properties must be easy for administrators to estimate, apply, and control. A possible solution could be a dashboard showing the aggregated re-authentication rates and risk scores per user. These metrics can help to control and adjust the %
thresholds continuously and whenever necessary.%

Even in settings involving a high TPR, the RBA models hardly ask for re-authentication at all. While this is a very good sign for the security properties of RBA, this influences users. Users will only feel protected by RBA if they get prompted for re-authentication at least once~\cite{wiefling_more_2020}. To support users in feeling protected, we suggest to inform about RBA being active%
.

\section{Analyzing RBA Features (RQ2)}\label{sec:features}

Based on our 247 collected features, we determined a subset %
that is suitable for RBA use. To be qualified for RBA use, we defined %
necessary criteria. The features need to:
\begin{enumerate*}[label=(\Alph*)]
	\item \textbf{Have both a good level of stability and at least minimum entropy}: In contrast to %
	fingerprinting properties %
	for tracking purposes~\cite{pugliese_long-term_2020}, we require a certain level of %
	entropy to make it harder for attackers to reproduce the feature values by simply brute forcing them. %
	This might cause RBA to ask for re-authentication at a higher frequency. However, showing RBA presence by very few re-authentication requests can lead to increased (perceived) security~\cite{wiefling_more_2020}.
	\item \textbf{Be spoofable only with a high amount of effort}: %
	Easy-to-guess features will not bring any attack detection advantage to the RBA feature set baseline.
    \item \textbf{Increase differentiation between legitimate users and attackers}: When added to the baseline feature set, %
    risk scores differences between legitimate users and attackers should increase.
\end{enumerate*}

\subsection{Study Setup}

Based on the defined criteria, we developed and conducted several big data computing jobs to analyze the performance of all features in our data set.

\subsubsection{Test~A: Entropy.}

To identify easy-to-spoof features, %
we calculated the Shannon entropy of the feature values $x_{i_j}$ of each feature $x_i \in X$ in the login history with $n = |x_{i}|$:
\begin{equation}
H_{x_i} = -\sum_{j=0}^{n} x_{i_j} \cdot log_2(x_{i_j})
\end{equation}

We calculated two variants of entropy. To observe overall differences, we calculated the entropy $H_{{global}_{x_i}}$ for the global login history. To observe the feature stability inside the login history of each user, we calculated the mean Shannon entropy $\overline{H_{{user}_{x_i}}}$ of each feature in the user's login history.
As a result, features with $H_{{global}_{x_i}} = 0$ did not contain any information to distinguish between users. Similarly, features with $\overline{H_{{user}_{x_i}}} = 0$ did not change inside the users' login histories.

\subsubsection{Test~B: Number of Feature Values.}

Some of the collected features can be spoofed by attackers with low effort. This is especially true for client submitted features, e.g., output of a JavaScript function executed in the user's browser.

To make %
features harder to guess for attackers, they need to have a large range of values with equal distribution%
. %
Assuming that %
accounts will be locked after RBA detected an illegitimate login%
, it will be difficult for attackers to guess correct feature values with increasing numbers of unique feature values.

\subsubsection{Test~C: Risk Score Changes.}

We studied the risk score behavior of the features to evaluate their potential to improve the detection of attackers and legitimate users. We tested the features with 
the EXTEND model since it provides fine grained risk scores. %

For each feature, we %
calculated the risk scores of all illegitimate login attempts by targeted attackers per user (attacker risk scores)%
. We then calculated the risk scores of all legitimate login attempts (legitimate risk scores). After that, we determined the risk score relation (RSR) as the relation between the mean attacker and mean legitimate risk scores:
\begin{equation}
    RSR_{basic} = \frac{mean\ attacker\ risk\ score}{mean\ legitimate\ risk\ score}
\end{equation}
To ease comparison%
, we normalized the RSRs for each feature $x_i \in X$ to the %
baseline:
\begin{equation}
    RSR_{x_i} = RSR_{basic_{x_i}} - RSR_{basic_{baseline}}
\end{equation}
The feature baseline varied depending on the feature being compared to, e.g., the IP address when all compared features were added to the IP address. When testing only one feature, the baseline was a feature without any entropy, to observe risk score differences when entropy was added.
If the RSR of a %
feature $x_i \in X$ is greater than the baseline RSR, i.e., $RSR_{x_i} > 0.0$, this feature increased the differentiation between legitimate users and attackers %
compared to the baseline.

\subsubsection{Subset Extraction.}

For each test, we defined the following thresholds to extract a subset of suitable features for RBA use:
\begin{enumerate*}[label=(Test \Alph*)]
\item To extract features having at least minimum entropy, we only considered features with $H_{{global}_{x_i}} > 0.1$ and $\overline{H_{{user}_{x_i}}} > 0.1$. Based on the third quantile and the specific characteristics of the data set, we chose this threshold as a minimum baseline.

\item To focus on harder-to-guess features for RBA, we considered those with more than ten unique feature values. More features were considered for both desktop and mobile users in the global login history to adequately address security. We made sure to check both mobile and desktop devices since mobile devices tend to have less unique RBA feature values than desktop devices~\cite{spooren_mobile_2015}.

\item To ignore features causing only small RSR improvements, we considered features with $RSR_{x_i} > 0.1$.
\end{enumerate*}

\subsubsection{Feature Reliability.}

The extracted features were present on all user sessions but were very diverse, %
ranging from client originated to server side recorded. Thus, we labeled them by the following properties%
:
\begin{enumerate*}
    \item \textbf{Server side}: These features are measured on the server side. Since they do not depend on client originated input%
    , they add a high level of trust.
    \item \textbf{Client side JavaScript not required}: There might be users that deactivated JavaScript, e.g., for privacy reasons. To ensure compatibility%
    , we labeled features that can be measured without JavaScript.
\end{enumerate*}

Based on the properties, we distinguished three categories of RBA features:
\textbf{Single features} add a high level of reliability and provide good RBA performance on their own. \textbf{Major add-on features} are similar, but they only achieve good RBA performance when added to a single feature. Both feature types can be used %
with high %
weighting and are measured on the server side.
\textbf{Add-on features} are not as reliable as %
the features above but can be used in addition to single features. %
They are client originated%
. Therefore, it is possible that some of them could %
be blocked or modified, e.g., by anti-tracking measures~\cite{bujlow_survey_2017}.

\subsubsection{Re-authentication Count Changes.}

We assume that less requests for re-authentication can increase RBA usability and user acceptance~\cite{wiefling_more_2020}. Thus, we measured whether certain features have the potential to decrease the %
requests for legitimate users. We calculated the median login count until re-authenti\-ca\-tion for average legitimate users (i.e., 12 logins) and a TPR of 0.8 (targeted attackers) for each feature. We selected the TPR to allow all features to get a TPR close to the desired TPR for fair comparison. Also, selecting targeted attackers allowed us to test the features against the best possible attacker.

High re-authentica\-tion counts can signal administrators to weigh this feature lower, in combination with other features having lower %
counts, to balance usability.

\subsection{Results}

In the following, we present our results ordered by the three RBA feature categories. For statistical testing, we used Kruskal-Wallis tests for the omnibus cases and Dunn's multiple comparison test with Bonferroni correction for post-hoc analysis. We considered p-values lower than 0.05 as significant.

We calculated the risk scores on a high-performance computing (HPC) cluster with more than 2400 CPU cores. This was necessary since such calculations were computationally intensive. %
Using the HPC cluster reduced the calculation time to approximately two days for all features (instead of 123.5 days using 32 cores).

\begin{table}[t]
    \centering
    \caption{Single and major add-on features that qualified for RBA use. The only single feature is the IP address (bold). The other ones are major features that can be used in addition to a single feature. All features are server originated and hence hard to spoof.}
    \resizebox{0.70\linewidth}{!}{%
    \begin{threeparttable}
    \begin{tabular}{@{}llrrrrr@{}}
        \toprule
        {} & JavaScript  &   &   &   &    & Median logins until \\
        Feature          &    not required    &  $RSR$ &  $H_{global}$ &  $\overline{H_{user}}$ & Unique values & re-authentication\\
        \midrule
        \textbf{IP address}                         &     \CIRCLE &  1.20 \progressbar{1.00}  &         10.51 &                   1.96 &  \CIRCLE\CIRCLE\CIRCLE\CIRCLE\CIRCLE &   **2.00 \progressbar[subdivisions=12]{0.55} \\
        \midrule
        RTT-10MS         &                 \Circle &  1.75 \progressbar{1.00}  &          2.45 &                   1.04 &  \CIRCLE\CIRCLE\Circle\Circle\Circle &  1.50 \progressbar[subdivisions=12]{0.36} \\
        RTT-5MS          &                 \Circle &  1.37 \progressbar{0.78}  &          3.27 &                   1.33 &  \CIRCLE\CIRCLE\Circle\Circle\Circle &  1.71 \progressbar[subdivisions=12]{0.45} \\
        ASN (IP)         &                 \CIRCLE &  0.91 \progressbar{0.52}  &          3.17 &                   0.76 &  \CIRCLE\CIRCLE\Circle\Circle\Circle &  3.00 \progressbar[subdivisions=12]{0.73} \\
        RTT-MS           &                 \Circle &  0.56 \progressbar{0.32}  &          5.43 &                   2.00 &  \CIRCLE\CIRCLE\CIRCLE\CIRCLE\Circle &  2.00 \progressbar[subdivisions=12]{0.55} \\
        Hour             &                 \CIRCLE &  0.23 \progressbar{0.13}  &          4.06 &                   2.31 &  \CIRCLE\Circle\Circle\Circle\Circle &  4.00 \progressbar[subdivisions=12]{0.82} \\
        Region (IP)      &                 \CIRCLE &  0.15 \progressbar{0.09}  &          1.20 &                   0.31 &  \CIRCLE\Circle\Circle\Circle\Circle &  1.71 \progressbar[subdivisions=12]{0.45} \\
        Weekday and hour &                 \CIRCLE &  0.15 \progressbar{0.08}  &          6.72 &                   2.78 &  \CIRCLE\CIRCLE\CIRCLE\Circle\Circle &  4.00 \progressbar[subdivisions=12]{0.82} \\
    \bottomrule
    \end{tabular}
    \begin{tablenotes}
    \small
    Significantly higher than the baseline:
    \item ** p $<$ 0.01 \\
    Baselines: Zero entropy feature (single feature), IP address (major add-on feature)\\
    Unique values: Five dot scale (very low, low, medium, high, very high) mapped to the values (10-24, 25-74, 75-149, 150-300, $>$300).\\
    \end{tablenotes}
    \end{threeparttable}
    }
    \vspace{-1.7em}
    \label{tab:results-single-feature}
    \label{tab:results-major-features}
\end{table}

After combining the features that passed all three tests, only the IP address qualified as a \textbf{single feature} for RBA use.
When being used in addition to the IP address, seven features qualified as \textbf{major add-on features}, all of them network or behavior based (see Table~\ref{tab:results-major-features}).
Since the IP address was the only appropriate single feature for this case, we extracted the \textbf{add-on features} using this feature. 27 %
features qualified by passing all three tests (see Table~\ref{tab:results-addon-features}).%

\begin{table}[t]
    \centering
    \caption{Add-on features that qualified for RBA use in addition to single features. In comparison to major add-on features, they are client originated and thus spoofable.}
    \resizebox{0.8\linewidth}{!}{%
    \begin{threeparttable}
    \begin{tabular}{@{}ll@{}rrrrr@{}}
    \toprule
    {} & JavaScript   &  &   &   &   & Median logins until \\
    Feature   &   not required      &  $RSR$ &  $H_{global}$ &  $\overline{H_{user}}$ &  Unique values & re-authentication \\
    \midrule
    Session Cookie                         &                 \CIRCLE &  22.39 \progressbar{6.82}  &          9.51 &                   0.51 &  \CIRCLE\CIRCLE\CIRCLE\CIRCLE\CIRCLE &  **12.00 \progressbar[subdivisions=12]{1.00} \\
    User agent string (w/ subfeatures) &                 \CIRCLE &  10.33 \progressbar{3.15}  &          7.43 &                   1.21 &  \CIRCLE\CIRCLE\CIRCLE\CIRCLE\CIRCLE &  **12.00 \progressbar[subdivisions=12]{1.00} \\
    Screen width and height            &                 \Circle &   3.28 \progressbar{1.00}  &          4.70 &                   0.64 &  \CIRCLE\CIRCLE\CIRCLE\CIRCLE\Circle &     3.00 \progressbar[subdivisions=12]{0.73} \\
    WebGL fingerprint                  &                 \Circle &   3.14 \progressbar{0.96}  &          4.12 &                   0.55 &  \CIRCLE\CIRCLE\CIRCLE\Circle\Circle &     4.00 \progressbar[subdivisions=12]{0.82} \\
    Accept language header             &                 \CIRCLE &   3.01 \progressbar{0.92}  &          2.57 &                   0.33 &  \CIRCLE\CIRCLE\CIRCLE\Circle\Circle &     3.00 \progressbar[subdivisions=12]{0.73} \\
    App version                        &                 \Circle &   2.74 \progressbar{0.83}  &          6.29 &                   1.01 &  \CIRCLE\CIRCLE\CIRCLE\CIRCLE\CIRCLE &     2.40 \progressbar[subdivisions=12]{0.64} \\
    Available width and height         &                 \Circle &   2.72 \progressbar{0.83}  &          6.36 &                   0.95 &  \CIRCLE\CIRCLE\CIRCLE\CIRCLE\CIRCLE &     3.00 \progressbar[subdivisions=12]{0.73} \\
    OS full version                    &                 \CIRCLE &   2.64 \progressbar{0.80}  &          3.90 &                   0.59 &  \CIRCLE\CIRCLE\CIRCLE\Circle\Circle &     2.40 \progressbar[subdivisions=12]{0.64} \\
    WebGL Version                      &                 \Circle &   2.59 \progressbar{0.79}  &          2.14 &                   0.36 &  \CIRCLE\CIRCLE\Circle\Circle\Circle &     2.40 \progressbar[subdivisions=12]{0.64} \\
    WebGL extensions                   &                 \Circle &   2.52 \progressbar{0.77}  &          3.62 &                   0.56 &  \CIRCLE\CIRCLE\Circle\Circle\Circle &     6.00 \progressbar[subdivisions=12]{0.91} \\
    HTML5 canvas fingerprint           &                 \Circle &   2.28 \progressbar{0.70}  &          6.45 &                   0.77 &  \CIRCLE\CIRCLE\CIRCLE\CIRCLE\CIRCLE &     3.00 \progressbar[subdivisions=12]{0.73} \\
    OS name and version                &                 \CIRCLE &   2.27 \progressbar{0.69}  &          3.91 &                   0.59 &  \CIRCLE\CIRCLE\CIRCLE\Circle\Circle &     2.40 \progressbar[subdivisions=12]{0.64} \\
    Browser major version              &                 \CIRCLE &   2.03 \progressbar{0.62}  &          4.28 &                   0.80 &  \CIRCLE\CIRCLE\Circle\Circle\Circle &   **6.00 \progressbar[subdivisions=12]{0.91} \\
    Device pixel ratio                 &                 \Circle &   1.94 \progressbar{0.59}  &          2.66 &                   0.51 &  \CIRCLE\CIRCLE\Circle\Circle\Circle &     2.40 \progressbar[subdivisions=12]{0.64} \\
    User agent string (no subfeatures) &                 \CIRCLE &   1.74 \progressbar{0.53}  &          7.43 &                   1.21 &  \CIRCLE\CIRCLE\CIRCLE\CIRCLE\CIRCLE &     3.00 \progressbar[subdivisions=12]{0.73} \\
    Main language                      &                 \Circle &   1.61 \progressbar{0.49}  &          1.35 &                   0.24 &  \CIRCLE\Circle\Circle\Circle\Circle &     3.00 \progressbar[subdivisions=12]{0.73} \\
    Browser full version               &                 \CIRCLE &   1.27 \progressbar{0.39}  &          5.49 &                   0.96 &  \CIRCLE\CIRCLE\CIRCLE\Circle\Circle &  **12.00 \progressbar[subdivisions=12]{1.00} \\
    Browser name and version           &                 \CIRCLE &   1.14 \progressbar{0.35}  &          5.85 &                   1.02 &  \CIRCLE\CIRCLE\CIRCLE\CIRCLE\Circle &   **6.00 \progressbar[subdivisions=12]{0.91} \\
    Local IP address                   &                 \Circle &   1.13 \progressbar{0.34}  &          3.27 &                   0.49 &  \CIRCLE\CIRCLE\CIRCLE\CIRCLE\CIRCLE &     1.00 \progressbar[subdivisions=12]{0.00} \\
    Webkit temporary storage           &                 \Circle &   0.92 \progressbar{0.28}  &          3.30 &                   0.39 &  \CIRCLE\CIRCLE\CIRCLE\CIRCLE\CIRCLE &     1.00 \progressbar[subdivisions=12]{0.00} \\
    Battery discharging time           &                 \Circle &   0.75 \progressbar{0.23}  &          1.95 &                   0.48 &  \CIRCLE\CIRCLE\CIRCLE\CIRCLE\CIRCLE &     1.00 \progressbar[subdivisions=12]{0.00} \\
    \bottomrule
    \end{tabular}
    \begin{tablenotes}
    \small
    Significantly higher than the baseline:
    \item ** p $<$ 0.01 \\
    Unique values: Five dot scale (very low, low, medium, high, very high) mapped to the values (10-24, 25-74, 75-149, 150-300, $>$300).\\
    The session cookie was set by the server. RBA simply compared the stored value.\\
    We omitted similar features for space reasons (see Table~\ref{tab:results-plus-ip-feature-appendix} in Appendix \ref{appendix:features} for all results).
    \end{tablenotes}
    \end{threeparttable}
    }
    \vspace{-1.3em}
    \label{tab:results-addon-features}
\end{table}

\subsubsection{Conclusion.}

In summary, a set of features has to be chosen in most cases rather than a single feature to achieve good RBA security. Using only one feature for RBA risk estimation will make it hard to reliably distinguish between attackers and legitimate users. The feature set needs to  at least include %
features that we identified as single or major add-ons (see Table~\ref{tab:results-major-features}) for good RBA security.%

\subsection{Discussion}
The results confirm \censorchange{our previous findings}{findings of Wiefling et al.}~\cite{wiefling_is_2019} that IP address, user agent%
, display resolution, language, and login time are useful RBA features and hence, find adoption in the wild. 
The results also show that most of the 247 analyzed features are not suitable for RBA use. Many of them had few unique values or low RSRs. This is good for privacy, as %
few features need to be collected. Also, many of the popular features~\cite{wiefling_is_2019} are collected on the server side anyway, e.g., in the %
logs~\cite{ibm_log_2003}. Still, some of them may contain sensitive data~\cite{bonneau_privacy_2014} and must be %
protected against data breaches. But, as we considered all features as categorical data%
, these can be hashed, or even truncated to some degree, to produce the same results%
.
Our results suggest a set of relevant RBA features may provide security benefits while preserving usability%
. %
This set is rather small compared to the 247 evaluated features. Thus, we discuss how to design %
a minimal RBA feature set to also balance privacy.
We discuss a selection of relevant features and feature combinations based on our results and findings in literature below%
.

\subsubsection{Features.}

The \textbf{IP address} proved to be the only RBA feature that can be used as a single feature. %
The \textbf{region} and 
\textbf{ASN} %
are also hard to fake and to obtain since they require network access from a specific ASN in a specific location. %

The \textbf{RTT} turned out as a promising new RBA feature when being rounded to milliseconds at least. %
Attackers %
need access to a device physically located inside the victim's location %
to forge this feature. Thus, using the RTT would add high costs for attackers%
. However, due to more re-authentication requests, the RTT needs to be weighted lower than other features to balance usability.

Timing features like \textbf{weekday and hour} increased security attributes while having few re-authentication requests. Successful attacks need to estimate the victim's usual login times right to the day and hour, which can be greater effort. This is especially the case for services that are not used %
on a daily basis.%

The \textbf{user agent string} performed very well when used in combinations with a subfeature hierarchy, confirming findings of Freeman et al.~\cite{freeman_who_2016}.%

Since it can be used as a unique %
session identifier, the \textbf{cookie} %
seems to be an obvious feature choice, and our results would support this view%
. However, cookies should only be used very carefully or not at all as a RBA feature. They would have to be stored permanently in the login history. Since there is no revocation mechanism in the current RBA models, every cookie inside the login history would always be valid%
.
Thus, a stolen and even outdated cookie might have a negative impact on the risk score, leading to false positives.

\subsubsection{Feature combinations.}

The \textbf{IP address and user agent string} features are often named in literature~\cite{freeman_who_2016,hurkala_architecture_2014,spooren_mobile_2015}. According to our observations related to the data set, they increased the RSR %
and significantly reduced re-authentications compared to %
the single features%
. %

RBA models in literature often use \textbf{user agent strings} to identify a browser \cite{hurkala_architecture_2014,spooren_mobile_2015,freeman_who_2016,wiefling_is_2019,djosic_machine_2020}. However, \textbf{HTML5 canvas and WebGL fingerprints}~\cite{mowery_pixel_2012,daud_adaptive_2017} are newer approaches considered more difficult to fake. %
Both approaches received lower RSRs and significantly higher re-authentication counts compared to the user agent string in our data set%
. %
Following that, %
if canvas or WebGL fingerprinting should be used to strengthen security, one should consider using them with lower weightings.

\section{Analyzing RBA Configuations (RQ3)}

For good usability, the latency between submitting the login credentials and getting the risk decision needs to be low. An acceptable delay ranges below 300~ms when considering the page load time~\cite{borzemski_impact_2018}. %
Thus, we analyzed which properties %
have an impact on the risk score calculation time. %
This can help to design RBA systems with both good security and a low authentication time.

We replayed all legitimate logins with both models and measured the time it took to calculate the risk score. %
We measured %
on a server with %
Intel Xeon Gold 6130 processor (2.1 GHz, 64 cores), 480 GB SSD storage, and 64 GB RAM. %
We used Kruskal-Wallis %
tests to check for significant differences between features. %
For variables suggesting a relation, we calculated the linear least squares regression between them. We determined the effect sizes based on Cohen~\cite{cohen_statistical_1988}.

\subsubsection{Test 1: Single Feature.}

We first measured the calculation times for every %
feature. %
The median calculation times %
ranged 4.5-8~ms for EXTEND (median: 5.63; SD: 0.9), and 0.07-2.7~ms for SIMPLE (median: 0.08; SD: 0.17). There were no significant differences between the features%
. However, there was a large significant effect between the calculation time and the global login history size for EXTEND%
. The linear regression yielded $y = 4.1912 + 0.0003\cdot x$, with $y$ being the time in ms and $x$ the global login history size ($R^2$=0.42; f=0.85; p$\ll$0.0001).

\subsubsection{Test 2: Adding Features.}

We measured the calculation time based on the number of features in the feature set. However, testing all $2^{247} - 1$ %
combinations was not feasible. Since there were no significant differences between all features in Test 1, we chose the feature that ranged in the middle of all median calculation times. We did this to select a feature that matches all features as well as possible.
We took this %
feature, added it to the feature set%
, measured the times, and did it again until we reached the maximum number of features found in RQ2%
.

\begin{figure}[t]
    \centering
    \includegraphics[width=0.97\linewidth]{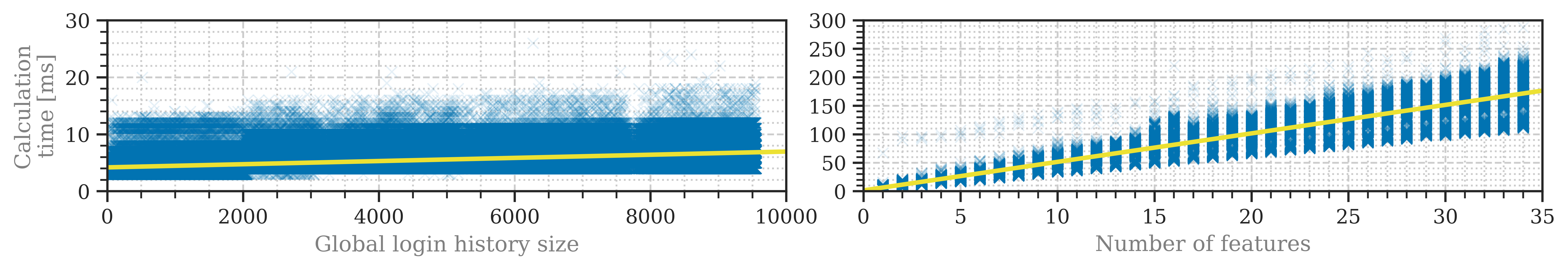}
    \caption{Relationship between risk score calculation time and the size of the global login history (left) or number of features (right) for EXTEND. The diagonal line represents the fitted linear regression model. Left: We limited the y-axis to 30~ms for readability.}
    \vspace{-1.3em}
    \label{fig:regression_time-global_history_size}
    \label{fig:regression_time-num_features}
\end{figure}

The results showed significant effects between the number of features %
and the calculation time (see Figure~\ref{fig:regression_time-num_features}). The fitted linear regression model resulted in $y=1.5568 + 5.0038\cdot x$ and a large effect size for EXTEND ($R^2$=0.93; f=3.71; p$\ll$0.0001), with $y$ being the %
time in ms and $x$ the number of features. Linear regression for SIMPLE resulted in $y=-0.0119 + 0.0013\cdot x$ and a medium effect ($R^2$=0.12; f=0.37; p$\ll$0.0001). However, the latter effects were hardly noticeable.

\subsubsection{Discussion.}

Administrators need to keep track of the included global login history and features to ensure an acceptable authentication speed. %
The results show that including a high amount of features %
impacts the performance for EXTEND. However, our results for RQ1 already showed that capturing few features was sufficient for good security and usability.

\section{Limitations}

We implemented the RBA models using Python. High-level programming languages like C++ might have reduced the calculation time. Nevertheless, our results can still give estimates on factors that influence RBA performance.

The results are limited to the data set tested and the users who participated. Our results are not representative of large-scale online services, but represent a typical use case scenario of a daily to weekly use online service in a certain country%
. We assume that the IP country feature would have qualified with an international user base~\cite{freeman_who_2016}.
To allow a fair comparison of all %
features, we weighted all features equally%
. We expect, however, that service owners weigh features individually, possibly improving the RBA performance. Thus, we assume that our study results represent a RBA performance baseline.

As in similar studies%
, we can never fully exclude that the website was targeted by intelligent attackers. However, we implemented multiple countermeasures%
.    
The website URL was only provided to %
students %
signing an informed consent. The URL was not accessible via %
search engines due to geoblocking and other measures to disallow %
crawling the site. %
IP scans reaching the website's IP address only received a white page instead of the e-learning website. The TLS certificate also did not reveal the real DNS entry %
in this case.
The fact that %
users did not notice illegitimate login attempts %
and %
no data breaches %
were known~\cite{haveibeenpwned_pwned_2020} underlines %
that the website was likely not infiltrated.

\section{Related Work}
 
In previous work, we studied RBA's usability characteristics~\cite{wiefling_more_2020,wiefling_evaluation_2020}. The results helped to estimate the usability of RBA characteristics in this study.
To the best of our knowledge, no studies analyzing RBA characteristics with long-term login data exist in literature.
Freeman et al.~\cite{freeman_who_2016} tested their RBA model on a LinkedIn data set using only IP address and user agent string as features. 
In contrast to them, we tested their model with a huge set of %
features.

There is also related work regarding browser fingerprinting features for user authentication purposes%
.
Alaca and van Oorschot~\cite{alaca_device_2016} classified 29 fingerprinting features which have the potential to be used for user authentication. They selected the features based on literature research but, in contrast to our study, did not test them on real data. %
Spooren et al.~\cite{spooren_mobile_2015} tested OpenAM's RBA mechanism on simulated data with six features, which were screen resolution, browser plugins, fonts, timezone, user agent, and geolocation. They found that mobile devices were less reliable in terms of being uniquely identified. We were able to confirm their findings for these six features. However, our study shows that there are other features that can reliably identify mobile device users.
Campobasso et al.~\cite{campobasso_impersonation_2020} studied a criminal infrastructure that tries to bypass RBA on malware infected victim devices. Since its geolocation spoofing relied on SOCKS5 proxies, our new RTT feature can detect these attacks.
Andriamilanto et al.~\cite{andriamilanto_guess_2021} tested fingerprints of website users regarding their capability to be used for authentication purposes. In contrast to our study, their data set did not relate to login attempts, contained only client originated features, and was not tested on RBA.

\section{Conclusion}
As long as password-based authentication predominates, constantly evolving data breaches and targeted attacks with breached passwords~\cite{akamai_credential_2019} increase the need of RBA for online services to protect their users. NIST recommends RBA use since 2017~\cite{grassi_digital_2017}. However, the current body of knowledge does not provide insights on RBA characteristics. Understanding these is important to ensure that practical RBA deployments protect users as much as possible while balancing usability. To close this gap, we studied RBA %
characteristics %
with long-term usage data of a real-world online service.
Our results show that RBA can achieve low re-authentication rates for legitimate users when blocking more than 99.45\% of targeted attacks with the EXTEND model. %
Moreover, our findings also show that only few of the 247 collected features can be considered useful for practical RBA deployments. The IP address is confirmed to be a must-have feature in general, but it should be enriched by add-on features. Among them, the introduced RTT showed to be a new promising feature. Cookies, however, should only be used with great care or not at all, as stolen credentials together with a stolen cookie might outweigh other features and falsely grant access.

Our contribution indicates that simply acquiring one of the commercially or freely available RBA solutions is not sufficient. They still need to be customized for the targeted online service in order to be optimized in terms of security and usability. We provided insights on how to select proper features, their weightings, and the access threshold. Based on our findings, we recommend to use RBA algorithms comparable to the introduced EXTEND model, since its security and usability properties outweighed the SIMPLE model. Overall, RBA protection should be put in place shortly after the first deployment, as the login history size did not affect it in our study.%

\censorhidden{\small
	\subsubsection{\small Acknowledgments.}
Thanks to the anonymous reviewers, our shepherd Gunes Acar, and Florian Dehling for their detailed feedback, which greatly helped improve the paper. We would like to thank Rudolf Berrendorf and Javed Razzaq for providing us a huge amount of computational power for our big data analysis. We also thank Gaston Pugliese for providing us his fingerprinting script, Annette Ricke and Jan Herrmann for their support and cooperation, and Tanvi Patil for proofreading the paper. This research was supported by the research training group ``Human Centered Systems Security'' (NERD.NRW) sponsored by the state of North Rhine-Westphalia. The Platform for Scientific Computing was supported by the German Ministry for Education and Research, and the Ministry for Culture and Science of the state North Rhine-Westphalia (research grant 13FH156IN6).
} 

	\bibliographystyle{splncs04}
	\bibliography{bibliography_full}
\appendix

\newpage

\section{Survey}\label{appendix:survey}

We balanced all survey questions where applicable to mitigate social desirability bias~\cite{shaeffer_comparing_2005}. The questions were presented in random order to randomly distribute ordering effects~\cite{kalton_effect_1982}. We varied the scale direction of the questions for a random half of survey participants. For questions without an ordinal scale, we randomized the response options for each participant. We did all this to randomly distribute response order bias~\cite{chan_response-order_1991,hartley_thoughts_2014}. We also included an attention check similar to previous work~\cite{wiefling_evaluation_2020} to improve data quality.

\subsection{Online Service}

Question~\ref{question:learning-support} and \ref{question:recommend} were on a five-point Likert scale including a ``don't know'' option.

\begin{enumerate}
	\setlength\itemsep{0em}
    \item Which of these online services did you use at least once in the last three years? \emph{[Multiple choice]}
    \begin{answerlist}
        \item \website
        \item Google
        \item Facebook
        \item Twitch
        \item \emph{[made-up online service that did not exist]}
        \item Other: \_\_\_\_\_\_\_
    \end{answerlist}
    
    \emph{The order of the subquestions varied randomly in this question.}
    
    \item How much or little did \website support you in learning the lecture material?\label{question:learning-support}
    
    (5 - Did fully support, 1 - Did not support at all)

    \item Please rate your agreement with the following statement:\\\textbf{I think I would recommend \website to other students.}\label{question:recommend}
    
    (5 - Strongly agree, 1 - Strongly disagree)
    
    \item As far as you know, has anyone ever illegitimately logged into your personal \website account?\footnote{Note that Shay et al.~\cite{shay_my_2014} used a different response option order in their paper. We aligned the options to ordinal order so that we could vary the scale direction for a randomly selected half of participants.}
    \begin{answerlist}
        \item Yes, more than once
        \item Yes, only once
        \item No
        \item I don't know
    \end{answerlist}
\end{enumerate}

\subsection{Demographics}

\begin{enumerate}
    \item How old are you?
    \begin{answerlist}
        \item 18-24
        \item 25-34
        \item 35-44
        \item 45-54
        \item 55-64
        \item 65-74
        \item 75 or older
        \item Prefer not to say
    \end{answerlist}
    
    \item What is your gender?
    \begin{answerlist}
        \item Female
        \item Male
        \item Non-Binary
        \item Prefer not to say
    \end{answerlist}
\end{enumerate}

\vfill

\newpage
\section{Features}\label{appendix:features}

\begin{table}[h]
    \centering
    \caption{List of single (bold) and (major) add-on features that qualified for RBA use. All features are present in all sessions of the data set.}

    \resizebox{0.9\linewidth}{!}{%
    \begin{threeparttable}
    \begin{tabular}{@{}l@{}rrrrrrrr@{}}
        \toprule
        & & & & & & & Median &  \\
        & & & & & & & logins &  \\
        {} & Server & JS not  &    &   &   &  Unique  & until &  \\
        Feature    &   side   &  required  & $RSR$ &  $H_{global}$ &  $\overline{H_{user}}$ &   values &  re-auth.  & p\\
        \midrule
        \textbf{IP address}                         &     \CIRCLE &                 \CIRCLE &   1.20 &         10.51 &                   1.96 &           4073 &                                **2.00 &  $<$0.0001 \\
        \midrule
        Session Cookie                             &     \Circle &                 \CIRCLE &  22.39 &          9.51 &                   0.51 &           1534 &                               **12.00 &  $<$0.0001 \\
        User agent string (w/ subfeatures) &     \Circle &                 \CIRCLE &  10.33 &          7.43 &                   1.21 &            638 &                               **12.00 &  $<$0.0001 \\
        Screen width and height            &     \Circle &                 \Circle &   3.28 &          4.70 &                   0.64 &            176 &                                  3.00 &          - \\
        WebGL fingerprint                  &     \Circle &                 \Circle &   3.14 &          4.12 &                   0.55 &             90 &                                  4.00 &     0.8057 \\
        Screen height                      &     \Circle &                 \Circle &   3.09 &          4.34 &                   0.64 &            126 &                                  4.00 &     0.3426 \\
        Accept language header             &     \Circle &                 \CIRCLE &   3.01 &          2.57 &                   0.33 &             91 &                                  3.00 &          - \\
        Available screen width             &     \Circle &                 \Circle &   2.93 &          4.38 &                   0.69 &            150 &                                  3.00 &          - \\
        Screen width                       &     \Circle &                 \Circle &   2.93 &          4.28 &                   0.63 &            138 &                                  4.00 &     0.8883 \\
        App version                        &     \Circle &                 \Circle &   2.74 &          6.29 &                   1.01 &            534 &                                  2.40 &          - \\
        Available width and height         &     \Circle &                 \Circle &   2.72 &          6.36 &                   0.95 &            411 &                                  3.00 &          - \\
        OS full version                    &     \Circle &                 \CIRCLE &   2.64 &          3.90 &                   0.59 &             93 &                                  2.40 &          - \\
        Available screen height            &     \Circle &                 \Circle &   2.59 &          5.91 &                   0.95 &            289 &                                  3.00 &          - \\
        WebGL Version                      &     \Circle &                 \Circle &   2.59 &          2.14 &                   0.36 &             56 &                                  2.40 &          - \\
        Supported languages                &     \Circle &                 \Circle &   2.53 &          2.54 &                   0.36 &             87 &                                  3.00 &          - \\
        WebGL extensions                   &     \Circle &                 \Circle &   2.52 &          3.62 &                   0.56 &             69 &                                  6.00 &     0.1601 \\
        HTML5 canvas fingerprint           &     \Circle &                 \Circle &   2.28 &          6.45 &                   0.77 &            386 &                                  3.00 &          - \\
        OS name and version                &     \Circle &                 \CIRCLE &   2.27 &          3.91 &                   0.59 &             95 &                                  2.40 &          - \\
        Browser major version              &     \Circle &                 \CIRCLE &   2.03 &          4.28 &                   0.80 &             57 &                                **6.00 &     0.0046 \\
        Device pixel ratio                 &     \Circle &                 \Circle &   1.94 &          2.66 &                   0.51 &             70 &                                  2.40 &          - \\
        RTT-10MS                           &     \CIRCLE &                 \Circle &   1.75 &          2.45 &                   1.04 &             51 &                                  1.50 &          - \\
        User agent string (no subfeatures) &     \Circle &                 \CIRCLE &   1.74 &          7.43 &                   1.21 &            635 &                                  3.00 &          - \\
        Main language                      &     \Circle &                 \Circle &   1.61 &          1.35 &                   0.24 &             21 &                                  3.00 &          - \\
        RTT-5MS                            &     \CIRCLE &                 \Circle &   1.37 &          3.27 &                   1.33 &             67 &                                  1.71 &          - \\
        Browser full version               &     \Circle &                 \CIRCLE &   1.27 &          5.49 &                   0.96 &            118 &                               **12.00 &     0.0005 \\
        Browser name and version           &     \Circle &                 \CIRCLE &   1.14 &          5.85 &                   1.02 &            161 &                                **6.00 &     0.0064 \\
        Local IP address                   &     \Circle &                 \Circle &   1.13 &          3.27 &                   0.49 &            716 &                                  1.00 &          - \\
        Webkit temporary storage           &     \Circle &                 \Circle &   0.92 &          3.30 &                   0.39 &            735 &                                  1.00 &          - \\
        ASN (IP)                           &     \CIRCLE &                 \CIRCLE &   0.91 &          3.17 &                   0.76 &             43 &                                  3.00 &          - \\
        Battery discharging time           &     \Circle &                 \Circle &   0.75 &          1.95 &                   0.48 &           1007 &                                  1.00 &     0.0860 \\
        Battery level                      &     \Circle &                 \Circle &   0.73 &          2.36 &                   0.75 &             99 &                                  1.00 &     0.0935 \\
        RTT-MS                             &     \CIRCLE &                 \Circle &   0.56 &          5.43 &                   2.00 &            170 &                                  2.00 &          - \\
        Hour                               &     \CIRCLE &                 \CIRCLE &   0.23 &          4.06 &                   2.31 &             24 &                                  4.00 &          - \\
        Region (IP)                        &     \CIRCLE &                 \CIRCLE &   0.15 &          1.20 &                   0.31 &             16 &                                  1.71 &          - \\
        Weekday and hour                   &     \CIRCLE &                 \CIRCLE &   0.15 &          6.72 &                   2.78 &            145 &                                  4.00 &     0.2117 \\
        \bottomrule
    \end{tabular}
    \begin{tablenotes}
    \small
    Significantly higher than the baseline:
    \item * p $<$ 0.05
    \item ** p $<$ 0.01 \\
    We omitted p-values of 1.0 for readability reasons.
    \end{tablenotes}
    \end{threeparttable}
    }
    \label{tab:results-plus-ip-feature-appendix}
\end{table}

\begin{table}
    \centering
    \caption{Overview of features that were captured in the study and included in the compiled data set}
    \resizebox{0.97\linewidth}{!}{%
\begin{threeparttable}
    \begin{tabular}{r@{\hskip 1em}lll@{\hskip 3em}r@{\hskip 1em}ll}
    \cmidrule[\heavyrulewidth]{1-3} \cmidrule[\heavyrulewidth]{5-7}
    \# & \textbf{Feature} &                      Example Value &        &              \#       &              \textbf{Feature} &                     Example Value \\
    \cmidrule{1-3} \cmidrule{5-7}
    1      &                       Accept &  application/json, text/ja... &       &       83 &                    Geolocation &                          True \\
    2      &               AcceptEncoding &             gzip, deflate, br &       &       84 &                     GetBattery &                         False \\
    3      &               AcceptLanguage &  en-US,en;q=0.9,es-MX;q=0.... &       &       85 &                  GlobalStorage &                         False \\
    4      &                      ActiveX &                         False &       &       86 &                         HasDNT &                   unspecified \\
    5      &                      Adblock &                         False &       &       87 &                  IeAddBehavior &                         False \\
    6      &                         Ajax &         XMLHttpRequest object &       &       88 &                      IndexedDB &                          True \\
    7      &                  AppCodeName &                       Mozilla &       &       89 &                             IP &                       8.8.8.8 \\
    8      &              AppMinorVersion &                             0 &       &       90 &                         IP\_ASN &                          1234 \\
    9      &                      AppName &                      Netscape &       &       91 &                        IP\_City &                   Los Angeles \\
    10     &                   AppVersion &  5.0 (Macintosh; Intel Mac... &       &       92 &                     IP\_Country &                           USA \\
    11     &            AudioChannelCount &                             2 &       &       93 &                       IP\_Local &                192.168.178.35 \\
    12     &        AudioChannelCountMode &                      explicit &       &       94 &                      IP\_Region &                    California \\
    13     &   AudioChannelInterpretation &                      speakers &       &       95 &                           IPv6 &  597f:5abd:410d:4c1:382f:f... \\
    14     &  AudioChannelMaxChannelCount &                             2 &       &       96 &                       IPv6\_ASN &                          1234 \\
    15     &   AudioChannelNumberOfInputs &                             1 &       &       97 &                       IsMobile &                         False \\
    16     &  AudioChannelNumberOfOutputs &                             0 &       &       98 &                           Java &                         False \\
    17     &                  AudioCtxWin &                          True &       &       99 &                       Language &                         en-US \\
    18     &             AudioDestination &                          True &       &      100 &                      Languages &            ['en-US', 'es-MX'] \\
    19     &     AudioMozAudioChannelType &                         False &       &      101 &                   LocalStorage &                          True \\
    20     &              AudioSampleRate &                         48000 &       &      102 &                          MathE &                  2.7182818285 \\
    21     &                   AudioState &                     suspended &       &      103 &                       MathLn10 &            2.3025850930000002 \\
    22     &                      Battery &                         False &       &      104 &                        MathLn2 &                  0.6931471806 \\
    23     &              BatteryCharging &                         False &       &      105 &                     MathLog10E &           0.43429448190000003 \\
    24     &          BatteryChargingTime &                             0 &       &      106 &                      MathLog2E &                  1.4426950409 \\
    25     &       BatteryDischargingTime &                          6967 &       &      107 &                         MathPi &                  3.1415926536 \\
    26     &                 BatteryLevel &                          0.74 &       &      108 &                     MathSqrt12 &                  0.7071067812 \\
    27     &              BrowserLanguage &                         en-US &       &      109 &                      MathSqrt2 &                  1.4142135624 \\
    28     &                 CacheControl &                      no-cache &       &      110 &                      MimeTypes &          \textasciitilde pdf\textasciitilde application/pdf \\
    29     &                       CnvsFP* &  data:image/png;base64,iVB... &       &      111 &                     MozBattery &                         False \\
    29     &                   CnvsFPHash &  3e03748c78fd4bd2bd9af855c... &       &      112 &                MozGetUserMedia &                          True \\
    30     &                  CnvsWinding &                          True &       &      113 &                       MsCrypto &                         False \\
    31     &                   ColorDepth &                            24 &       &      114 &                         OnLine &                          True \\
    32     &                   Connection &                         close &       &      115 &                    OpenDBdroid &                         False \\
    33     &                ContentLength &                         47309 &       &      116 &                      OpenDBnav &                         False \\
    34     &                  ContentType &  application/json; charset... &       &      117 &                      OpenDBwin &                          True \\
    35     &                       Cookie &  SESSION="2e9d42f34b09adae... &       &      118 &                         Origin &           https://example.com \\
    36     &               CookiesEnabled &                          True &       &      119 &                          Oscpu &  Windows NT 10.0; Win64; x... \\
    37     &                          Cpu &                         empty &       &      120 &                     PixelDepth &                            24 \\
    38     &                       Crypto &                          True &       &      121 &                       Platform &                      MacIntel \\
    39     &                  Device\_name &              Samsung SM-G950A &       &      122 &                        Plugins &  \{'desc': '', 'file': '', ... \\
    40     &             DevicePixelRatio &                           2.0 &       &      123 &                         Pragma &                      no-cache \\
    41     &                          DNT &                           1.0 &       &      124 &                         Reader &             Chrome PDF Viewer \\
    42     &                     DntMSDNT &                     noSupport &       &      125 &                        Referer &  https://example.com/subur... \\
    43     &                       DntNav &                             1 &       &      126 &                       RTT-10MS &                          30.0 \\
    44     &                       DntWin &                             1 &       &      127 &                        RTT-5MS &                            35 \\
    45     &                       DotNet &  2.0.50727;3.5.30729;3.0.3... &       &      128 &               RTT-Measurements &  [22.551, 36.875, 31.619, ... \\
    46     &                        Flash &                26.0.0-blocked &       &      129 &                         RTT-MS &                            28 \\
    47     &            Flash32BitSupport &                     noSupport &       &      130 &                        RTT-RAW &                         29.95 \\
    48     &            Flash64BitSupport &                     noSupport &       &      131 &                       SaveData &                            on \\
    49     &       FlashAVHardwareDisable &                     noSupport &       &      132 &       Screen\_AvailWidth\_Height &                      834x1194 \\
    50     &         FlashcpuArchitecture &                     noSupport &       &      133 &            Screen\_Width\_Height &                      1280x720 \\
    51     &        FlashHasAccessibility &                     noSupport &       &      134 &              ScreenAvailHeight &                           824 \\
    52     &         FlashHasAudioEncoder &                     noSupport &       &      135 &               ScreenAvailWidth &                          1280 \\
    53     &        FlashHasEmbeddedVideo &                     noSupport &       &      136 &                   ScreenHeight &                           667 \\
    54     &                  FlashHasIME &                     noSupport &       &      137 &                    ScreenWidth &                          1280 \\
    55     &                  FlashHasMP3 &                     noSupport &       &      138 &                   Secfetchdest &                         empty \\
    56     &             FlashHasPrinting &                     noSupport &       &      139 &                   SecFetchMode &                          cors \\
    57     &      FlashHasScreenBroadcast &                     noSupport &       &      140 &                   SecFetchSite &                   same-origin \\
    58     &       FlashHasScreenPlayback &                     noSupport &       &      141 &                 SecurityPolicy &                     noSupport \\
    59     &       FlashHasStreamingAudio &                     noSupport &       &      142 &                 SessionStorage &                          True \\
    60     &       FlashHasStreamingVideo &                     noSupport &       &      143 &          SilverlightComponents &  1.0;2.0.30226;2.0.30523;2... \\
    61     &                  FlashHasTLS &                     noSupport &       &      144 &                 SystemLanguage &                         en-US \\
    62     &         FlashHasVideoEncoder &                     noSupport &       &      145 &                      Timestamp &           2018-10-15 22:17:35 \\
    63     &                    FlashLang &                     noSupport &       &      146 &                 Timestamp\_hour &                            10 \\
    64     &    FlashLocalFileReadDisable &                     noSupport &       &      147 &              Timestamp\_weekday &                             2 \\
    65     &            FlashManufacturer &                     noSupport &       &      148 &         Timestamp\_weekday\_hour &                           212 \\
    66     &             FlashMaxLevelIDC &                     noSupport &       &      149 &                 TimezoneOffset &                           480 \\
    67     &                      FlashOS &                     noSupport &       &      150 &                      UserAgent &  Mozilla/5.0 (Windows NT 1... \\
    68     &              FlashPlayerType &                     noSupport &       &      151 &           UserAgentBrowserName &                          Edge \\
    69     &             FlashScreenColor &                     noSupport &       &      152 &   UserAgentBrowserName\_version &              Chrome 73.0.3683 \\
    70     &               FlashScreenDPI &                     noSupport &       &      153 &   UserAgentBrowserVersion\_full &                      17.17134 \\
    71     &            FlashScreenHeight &                     noSupport &       &      154 &  UserAgentBrowserVersion\_major &                            12 \\
    72     &             FlashScreenWidth &                     noSupport &       &      155 &           UserAgentDeviceBrand &                         Apple \\
    73     &  FlashStageBrowserZoomFactor &                     noSupport &       &      156 &           UserAgentDeviceModel &                        iPhone \\
    74     &   FlashStageFullScreenHeight &                     noSupport &       &      157 &            UserAgentDeviceType &                        mobile \\
    75     &    FlashStageFullScreenWidth &                     noSupport &       &      158 &               UserAgentOS\_name &                      Mac OS X \\
    76     &         FlashTouchscreenType &                     noSupport &       &      159 &       UserAgentOS\_name\_version &                    Windows 10 \\
    77     &                 FlashVersion &                     noSupport &       &      160 &       UserAgentOS\_version\_full &                            10 \\
    78     &                      Fmradio &                         False &       &      161 &      UserAgentOS\_version\_major &                             5 \\
    79     &                        Fonts &  Agency FB; Aharoni; Alger... &       &      162 &                   UserLanguage &                         en-US \\
    80     &               FontsSmoothing &                          True &       &      163 &                         Vendor &          Apple Computer, Inc. \\
    81     &                     FontsSrc &                            js &       &      164 &                      VendorSub &                         empty \\
    82     &                        Gears &                     noSupport &       &      165 &                        Vibrate &                         False \\
    \cmidrule[\heavyrulewidth]{1-3} \cmidrule[\heavyrulewidth]{5-7}
    \end{tabular}
    \begin{tablenotes}
    \small
    \item * Mentioned for completeness. We used the collision free hash values in our studies. The canvas fingerprint is based on the FingerprintJS algorithm.
    \end{tablenotes}
\end{threeparttable}
    }
    \label{tab:feature-overview}
\end{table}

\begin{table}
\ContinuedFloat
    \centering
    \caption{Overview of features that were captured in the study and included in the compiled data set (continued)}
    \resizebox{0.63\linewidth}{!}{%
    \begin{threeparttable}

        \begin{tabular}{r@{\hskip 1em}ll}
        \toprule
        \# & \textbf{Feature} &                      Example Value  \\
        \midrule
        166    &                       WebGLaliasedLineWidthRange &                        [1, 1] \\
        167    &                       WebGLaliasedPointSizeRange &                     [1, 1024] \\
        168    &                                   WebGLalphaBits &                           8.0 \\
        169    &                                WebGLantiAliasing &                           yes \\
        170    &                                    WebGLblueBits &                           8.0 \\
        171    &                                   WebGLdepthBits &                          24.0 \\
        172    &                                  WebGLextensions &  EXT\_blend\_minmax; EXT\_sRG... \\
        173    &                                          WebGLFP* &  data:image/png;base64,iVB... \\
        173    &                                      WebGLFPHash &  4a5aab6eaab1cfe4f3eb8e90a... \\
        174    &            WebGLfragmentShaderHighFloatPrecision &                          23.0 \\
        175    &    WebGLfragmentShaderHighFloatPrecisionRangeMax &                           127 \\
        176    &    WebGLfragmentShaderHighFloatPrecisionRangeMin &                         127.0 \\
        177    &              WebGLfragmentShaderHighIntPrecision &                           0.0 \\
        178    &      WebGLfragmentShaderHighIntPrecisionRangeMax &                          30.0 \\
        179    &      WebGLfragmentShaderHighIntPrecisionRangeMin &                          31.0 \\
        180    &             WebGLfragmentShaderLowFloatPrecision &                          23.0 \\
        181    &     WebGLfragmentShaderLowFloatPrecisionRangeMax &                         127.0 \\
        182    &     WebGLfragmentShaderLowFloatPrecisionRangeMin &                         127.0 \\
        183    &               WebGLfragmentShaderLowIntPrecision &                           0.0 \\
        184    &       WebGLfragmentShaderLowIntPrecisionRangeMax &                          30.0 \\
        185    &       WebGLfragmentShaderLowIntPrecisionRangeMin &                          31.0 \\
        186    &          WebGLfragmentShaderMediumFloatPrecision &                          23.0 \\
        187    &  WebGLfragmentShaderMediumFloatPrecisionRangeMax &                          15.0 \\
        188    &  WebGLfragmentShaderMediumFloatPrecisionRangeMin &                         127.0 \\
        189    &            WebGLfragmentShaderMediumIntPrecision &                           0.0 \\
        190    &    WebGLfragmentShaderMediumIntPrecisionRangeMax &                          30.0 \\
        191    &    WebGLfragmentShaderMediumIntPrecisionRangeMin &                          31.0 \\
        192    &                                   WebGLgreenBits &                           8.0 \\
        193    &                    WebGLhasShaderPrecisionFormat &                          True \\
        194    &                               WebGLmaxAnisotropy &                          16.0 \\
        195    &                WebGLmaxCombinedTextureImageUnits &                           8.0 \\
        196    &                       WebGLmaxCubeMapTextureSize &                       16384.0 \\
        197    &                   WebGLmaxFragmentUniformVectors &                        1024.0 \\
        198    &                         WebGLmaxRenderBufferSize &                       16384.0 \\
        199    &                        WebGLmaxTextureImageUnits &                          16.0 \\
        200    &                              WebGLmaxTextureSize &                        4096.0 \\
        201    &                           WebGLmaxVaryingVectors &                          30.0 \\
        202    &                            WebGLmaxVertexAttribs &                          16.0 \\
        203    &                  WebGLmaxVertexTextureImageUnits &                           8.0 \\
        204    &                     WebGLmaxVertexUniformVectors &                         128.0 \\
        205    &                             WebGLmaxViewportDims &                [32767, 32767] \\
        206    &                                     WebGLredBits &                           8.0 \\
        207    &                                    WebGLRenderer &                  WebKit WebGL \\
        208    &                      WebGLShadingLanguageVersion &  WebGL GLSL ES 1.0 (OpenGL... \\
        209    &                                 WebGLStencilBits &                           0.0 \\
        210    &                                      WebGLVendor &                        WebKit \\
        211    &                                     WebGLVersion &  WebGL 1.0 (OpenGL ES 2.0 ... \\
        212    &              WebGLvertexShaderHighFloatPrecision &                          23.0 \\
        213    &      WebGLvertexShaderHighFloatPrecisionRangeMax &                         127.0 \\
        214    &      WebGLvertexShaderHighFloatPrecisionRangeMin &                           127 \\
        215    &                WebGLvertexShaderHighIntPrecision &                           0.0 \\
        216    &        WebGLvertexShaderHighIntPrecisionRangeMax &                          30.0 \\
        217    &        WebGLvertexShaderHighIntPrecisionRangeMin &                          24.0 \\
        218    &               WebGLvertexShaderLowFloatPrecision &                          10.0 \\
        219    &       WebGLvertexShaderLowFloatPrecisionRangeMax &                         127.0 \\
        220    &       WebGLvertexShaderLowFloatPrecisionRangeMin &                         127.0 \\
        221    &                 WebGLvertexShaderLowIntPrecision &                           0.0 \\
        222    &         WebGLvertexShaderLowIntPrecisionRangeMax &                          30.0 \\
        223    &         WebGLvertexShaderLowIntPrecisionRangeMin &                          24.0 \\
        224    &            WebGLvertexShaderMediumFloatPrecision &                          23.0 \\
        225    &    WebGLvertexShaderMediumFloatPrecisionRangeMax &                         127.0 \\
        226    &    WebGLvertexShaderMediumFloatPrecisionRangeMin &                         127.0 \\
        227    &              WebGLvertexShaderMediumIntPrecision &                           0.0 \\
        228    &      WebGLvertexShaderMediumIntPrecisionRangeMax &                          24.0 \\
        229    &      WebGLvertexShaderMediumIntPrecisionRangeMin &                          31.0 \\
        230    &                                    WebkitBattery &                         False \\
        231    &                               WebkitGetUserMedia &                          True \\
        232    &                           WebkitTemporaryStorage &                 3065883107320 \\
        233    &                                           WebRTC &                          True \\
        234    &                                      WebRTCAudio &                         False \\
        235    &                                 WebRTCDeviceEnum &                         False \\
        236    &                                     WebRTCdevIds &                         False \\
        237    &                                        WebRTCmoz &                          True \\
        238    &                             WebRTCRTPDataChannel &                         False \\
        239    &                            WebRTCScreenCapturing &                         False \\
        240    &                            WebRTCSCTPDataChannel &                          True \\
        241    &                                      WebRTCVideo &                         False \\
        242    &                                     WebRTCwebkit &                          True \\
        243    &                                   XHolaRequestId &                       96427.0 \\
        244    &                               XHolaUnblockerBext &  reqid 15019: before reque... \\
        245    &                                   Xjpoiaheoihgae &  eyJ0b2tlbiI6Iml0LWlzLW5vd... \\
        246    &                                   XRequestedWith &                XMLHttpRequest \\
        247    &                                  XTKAURLProtocol &                               \\
        \bottomrule
        \end{tabular}
        \begin{tablenotes}
        \small
        \item * Mentioned for completeness. We used the collision free hash values in our studies.
        \end{tablenotes}
    \end{threeparttable}
    }
    \label{tab:feature-overview2}
\end{table} \end{document}